\documentclass{aastex63}

\submitjournal{AJ}

\shorttitle{Switchback SEP}
\shortauthors{McDougall and Poduval}

\usepackage{xcolor}   
\usepackage{gensymb}

\usepackage{soul}
\usepackage{verbatim}
\usepackage{textcomp}
\usepackage{wasysym}
\newcommand{\isis}{IS{\astrosun}IS}
\usepackage{float}
\usepackage[caption = false]{subfig}
\usepackage{paralist}
\usepackage{rotating}

\usepackage{appendix}
\usepackage{longtable}
\newcommand{\kms}{\text{km}~\text{s}^{-1}}

\usepackage[color=red!20!white,textsize=tiny]{todonotes}

\begin{document}
  \title{Energetic Ion Composition as a Means of Investigating the Physical 
          Origins of Alpha Particle Heavy Magnetic Switchbacks}

\author[1234-1234-1234-1234]{Emily McDougall}
\affiliation{University of New Hampshire, Durham, NH}, 

\author[1234-1234-1234-1234]{Bala Poduval}
\affiliation{University of New Hampshire, Durham, NH}

\correspondingauthor{Emily McDougall}
\email{Emily.McDougall@unh.edu}


\begin{abstract}
Magnetic switchbacks are of continuing interest to the scientific community because the phenomenon has not been completely understood. Although most of the research into them in the Parker Solar Probe era has largely focused on creating a theoretical framework for causing the field reversal through magnetic interchange reconnection, reconnecting streams of plasma in the solar wind, or shear driven turbulence, it remains unclear to what extent these models may or may not represent the underlying physical reality of magnetic switchbacks. In this paper, we present the results of our study on the energetic ion composition of magnetic switchback events using statistical methods with the aim of obtaining new insights into the underlying physics.  In doing so, we find evidence that switchbacks containing suprathermal alpha flux display traits that are indicative of the Zank model of a magnetic 
"kink" propagating as a fast magnetosonic wave. These switchbacks also correlated with increased solar activity such as CMEs, high-speed streams, and solar flares. At the same time, switchbacks not meeting this criteria appear to occur at rates that are unaffected by increasing solar activity which implies the existence of multiple populations of switchbacks with differing physical mechanisms behind them. 

\end{abstract}

\section{Introduction}

Magnetic Switchbacks are quick reversals of the interplanetary magnetic field directions associated with sharp velocity spikes in the solar wind. Although magnetic switchbacks (switchbacks, hereafter) do not have a characteristic scale or duration, they exhibit several characteristic features such as low variability of the total magnetic field \citep{Larosa2021}, a difference in the solar wind speed relative to the outside plasma \citep{Kasper2019}, a reversal of the electron pitch angles inside the switchback, and an increase or decrease in the proton density relative to the outside plasma \citep{Huang2023}. Switchbacks were first detected and described, albeit not under the name magnetic switchbacks, based on observations in the Ulysses data \citep{Balogh1999,Goldstein1995} at 1.35 AU. They were subsequently understood to have been previously observed in Helios~1 and 2 data \citep{Borovsky2016, Horbury2018, Mariani1979} and were also subsequently observed during the first encounter with the solar corona of Parker Solar Probe (PSP) in November 2018 \citep{Bale2019, Kasper2019} and subsequent encounters \citep{Bowen2020, McComas2019, Schwadron2021, DudockdeWit2020, Horbury2020, Mozer2020, Rouillard2020, Tenerani2020}. In total, switchbacks have been observed by various probes at distances ranging from 0.2 AU to 2.9 AU (AU; the mean Sun-earth distance). 

\begin{figure}[htbp]
  \centering
  \centerline{
  \includegraphics[width=0.9\textwidth,clip=]{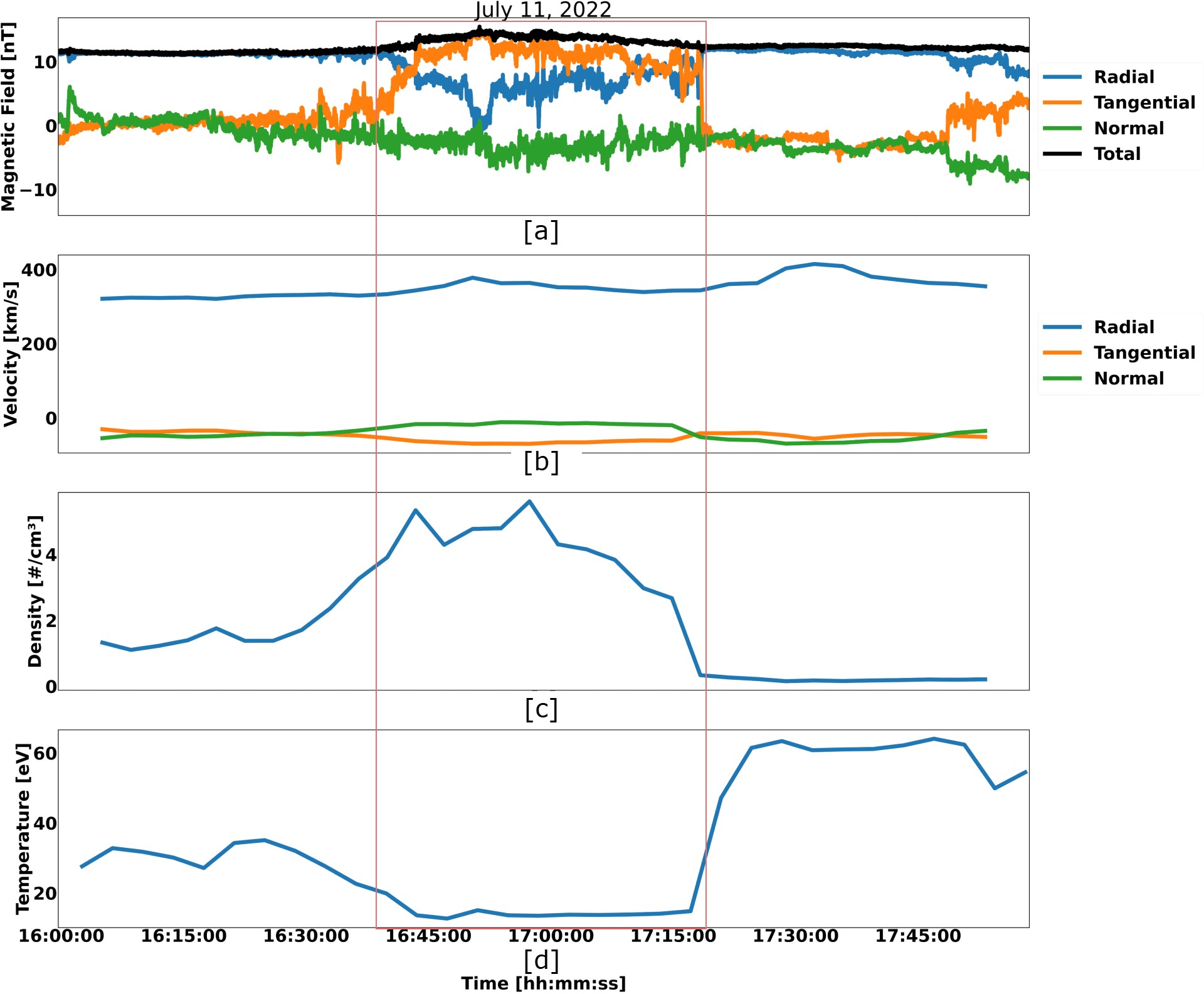}}
  \caption{  Plot of a switchback where the radial component of the solar wind velocity extends beyond the borders of the switchback. [a]: FIELDS magnetic field data. [b]: SWEAP solar wind velocity in RTN coordinates. [c]: SWEAP density data . [d]: SWEAP temperature data. The switchback is highlighted within a red box. } 
          
          \label{fig:Switchback}
\end{figure}

Figure \ref{fig:Switchback} shows an example of a switchback measured by the Parker Solar Probe at 0.74 AU. Panel~[a] shows the total magnetic field and three components of the magnetic field in RTN coordinates \citep{Farrugia2012}, where R points radially from Earth to the point of observation, N is normal to the ecliptic, and T points tangentially completing the right-handed coordinate system. The magnetic field components are seen to rotate starting at 16:30 and then “switch back” to their near-original orientation 45 minutes later (hence the name switchback). The velocity (Panel~(b)) changes over the same period with an increase in the radial and normal directions and a decrease in the tangential direction. However, the proton density (Panel~(c)) increases from 16:50 to 17:15 when the T and N components reverse and switches back, while the radial component decreases before increasing again at 17:30. The temperature also shows signatures over the same period of the switchback field reversal. 

\begin{figure}[htbp]
  \centering
  \centerline{
  \includegraphics[width=0.9\textwidth,clip=]{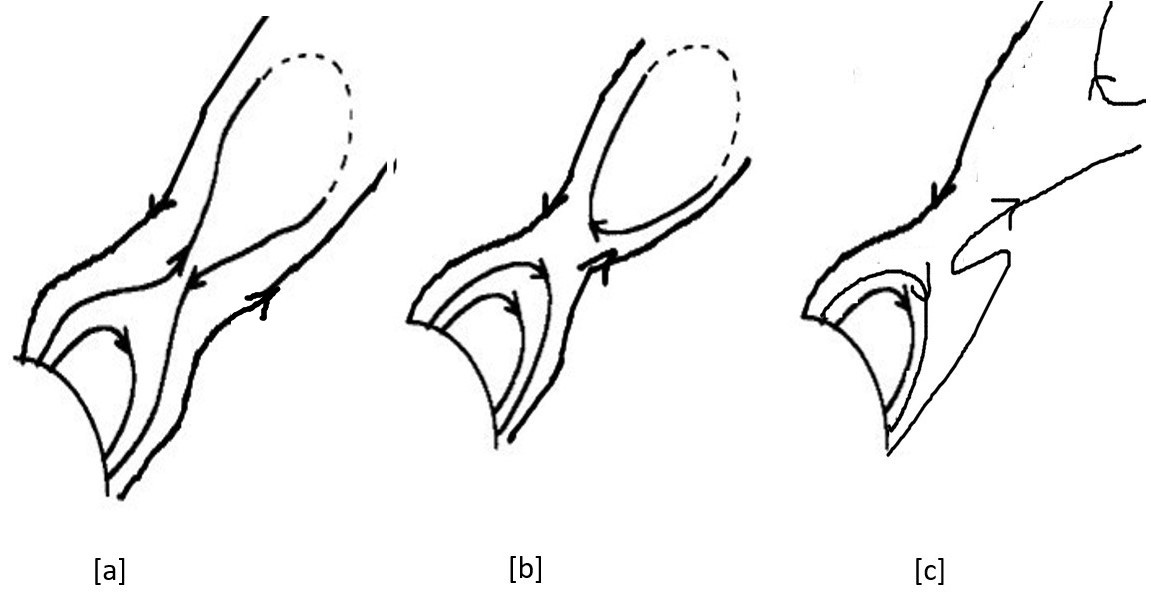}}
  \caption{ Diagram of a reconnection process where a magnetic "kink" might form in the wake of a CME flux rope moving out of the corona. Adapted from \cite{Crooker2002}.  
          } 
          \label{fig:Model}
\end{figure}

Several models have been put forth by various research groups to explain the origin of switchbacks although a consensus as to which one is the best fit for the phenomena has yet to be reached. Two of the models involve magnetic interchange reconnection of which the linear model or the Zank model \cite{Zank2020} is the simplest. It may occur with a single open and single closed field line, resulting in a linear model of interchange reconnection where the open and closed field lines reconnect into an “S” shape kink held together as it propagates as a fast magnetosonic wave. A fast magnetosonic wave is a normal mode of MHD, and normal modes such as these propagate with a phase velocity of 
\begin{equation} v_{\pm}^2 = \frac{1}{2} \left( c_{ms}^2 \pm \sqrt{c_{ms}^4 - 4 v_A^2 c_s^2 \cos^2 \theta} \right)
\label{eq:1}
\end{equation}
where $v_A$ is the Alfv\'{e}n velocity, $\theta$ is the angle between the equilibrium magnetic field $B_{0}$ and the wavevector $k$, $c_{s}$ is the speed of sound in the plasma, and $c_{ms}$ is the speed of propagation of a fast magnetosonic wave where
\begin{equation} 
c_{ms} = \sqrt{v_A^2 + c_s^2} 
\label{eq:2}
\end{equation} 
\citep{Zank2020}. A cartoon depicting how such a process may occur in the wake of the flux rope of a Coronal Mass Ejection (CME) extending from the photosphere into the corona is depicted in Figure \ref{fig:Model}. Similar interactions might exist in active regions with similar geometries, meaning that a similar process could also occur in solar flares, or high-speed streams \citep{Nandy2021, Yazev2023}. On the other hand, the other model involving interchange reconnection suggests that switchbacks may not be an “s” shaped kink at all, but it may instead be a flux rope formed in the interchange reconnection processes. This flux rope model originates from the study conducted by \cite{Neugebauer2012} that notes that polar X-ray jets were the sources of velocity peaks associated with microstreams in the high-speed component of the solar wind. Subsequent studies by \cite[e.g.][]{SterlingandMoore2015,Neugebauer2021} links such X-ray jets as a potential source of switchbacks and noted that the high correlation of switchbacks with microstreams \citep{Neugebauer1995} meant that a possible source for switchbacks was the result of eruptions of small-scale magnetic structures called mini-filaments into a high-speed jet of plasma from the solar corona.

Another prevailing theory of switchback formation is based on flux ropes. The \cite{Drake2021} model expands upon the geometry in Figure \ref{fig:Model} and notes that open field lines can exert magnetic pressure on a cylindrical flux bundle extending from the solar surface. The merging of open and closed field lines will develop a current sheet along the boundary of the cylindrical flux bundle. This current sheet would subsequently thin out and develop as a detached flux rope-type structure as the open and closed field lines reconnect \citep{Drake2021} which will contain the plasma that once made up the current sheet as it propagates. \cite{Drake2021} proposes that these resulting flux ropes are switchbacks entraining plasma from the solar corona. This entraining occurs because the flux rope once initially formed will relax to a state in which the plasma velocity flow is aligned with the magnetic field direction and remain constant within the flux surface that defines the flux rope. This relaxation would occur due to perpendicular electric fields, which are required to produce flows perpendicular to the magnetic field. These perpendicular electric fields would decay as the plasma reaches a more stable equilibrium state. The decay of perpendicular electric fields is also a process that often happens at the edge of flux rope structures due to reconnection \citep{Huang2008}, so it stands to reason that a stable flux rope model switchback would show consistent evidence of reconnection along its edges.

Other models explaining switchbacks are based on the dynamics of the solar wind as it propagates outward from the Sun. One such model suggests that switchbacks simply form as small perturbations in the solar plasma expand in the solar wind as they propagate away from the Sun \citep{Squire2020}. Another model known as the Alfv\'{e}nic turbulence model presents switchbacks as a form of Alfv\'{e}n waves caused by solar wind shear \citep{Landi2006, Ruffolo2020,Tenerani2020}. This model was created to explain the Alfv\'{e}nic properties of the switchback plasma and to explain the correlation between the time evolution of the plasma velocity and the group velocity of Alfv\'{e}n waves in switchbacks. A combined model presumes a combination of both interchange reconnection and shear flow, where interchange reconnection occurs in the solar corona, but is distorted via shear flow in the plasma structure as it propagates outward into the solar wind \citep{Schwadron2021}. Finally, \cite{Phan2020} proposed switchbacks are incidents of the spacecraft crossing of the heliospheric current sheet.

\cite{DudockdeWit2020} concluded that the quiescent solar wind (a region of the solar wind with low amplitude Alfv\'{e}nic fluctuations) and associated “memory-less” turbulence, which is a regime where the switchbacks distribution is statistically distinct from the turbulence, originates from low in the solar corona, “well below the Alfv\'{e}n surface”. This is hard to reconcile with models of switchbacks that presume the origin of switchbacks lies in the solar wind, such as the heliospheric current sheet model.

There are reasons to think that magnetic reconnection may play a part in at least some switchbacks. Firstly, because thermal ions in the solar wind are typically in the range of 0.5 to 10~keV, suprathermal particles which are more energetic than that in the solar wind would be associated with further methods of accelerating ions, such as magnetic reconnection. Ions above 20~keV/nucleon in-particular are associated with magnetic reconnection \citep{Reames1997, Jain2024} and have been observed in switchbacks \citep{Phan2020}. Secondly, many switchbacks do not display a reversal of direction of the electron strahl with respect to the local magnetic field, which is hard to explain via a model involving crossing the heliospheric current sheet but would not contradict a model involving magnetic reconnection \citep{Kasper2019}. Thirdly, it is easier to explain the observed increase in the ion temperature within many switchbacks relative to the surrounding plasma \citep{Farrell2020, Mozer2020} with reconnection than with Alfv\'{e}nic turbulence, given that non-resonant turbulence tends to preserve the apparent temperature of the plasma \citep{Nariyuki2012}. Fourthly, switchbacks often exhibit both magnetic pressure $(B_{P})$  and magnetic normal and tangential component $(B_{N,T})$ variation as linked to the presence of a transverse bulk flow of around 20 $\kms$ in the heliospheric azimuthal (T direction) near perihelion \citep{Kasper2019}. This is not indicative of Alfv\'{e}nic waves (which have constant $B_{P}$), but likely magnetosonic waves because $B_{P}$ variation is the defining signature of fast magnetosonic modes e.g., \cite{Lighthill1960}. This observation is far more consistent with \cite{FiskandKasper2020}, who suggest that there is a general azimuthal circulation of magnetic flux and plasma flow due to interchange reconnection in the low corona that could partially help explain magnetic switchbacks. Fifthly, the fact that magnetic switchbacks have been known to increase their ion temperature relative to the surrounding plasma. This has been known to exist within magnetic reconnection events, but is hard to explain via turbulence or solar wind flow \citep{Gosling2007,Drake2009}. 

Despite this, as to date there has not been much work done with the aim of empirically supporting one particular model or another. Thus, the goal of this study is to further refine the scientific understanding of the physics of switchbacks beyond a collection of possible models. For this, we will use the Parker Solar Probe ($PSP$) data to analyze specific parameters within switchbacks and demonstrate that the properties the switchbacks display are most indicative of interchange reconnection driven fast magnetosonic waves.

The paper is organized as follows: \S~\ref{sec:datamethod} describes the data used and the methodology we adopted, \S~\ref{sec:results} discusses the results of our methodology and introduces the plots and tables containing the data, \S~\ref{sec:discussion} discusses the implications of the data, and \S~\ref{sec:conclusion} recaps the end conclusions we can take away from this study. 
%
%
\section{Data and Methodology} \label{sec:datamethod}

\subsection{Data} \label{sub:data}

Parker Solar Probe was launched on August 12, 2018. The probe can reach up to a distance of 9.86 at its closest approach \citep{Fox2016}, making in-situ observations of the solar corona and solar wind. For the present study on switchbacks, we used data from FIELDS \citep{Bale2016}, the Integrated Science Investigation of the Sun \citep[\isis:][]{McComas2016,McComas2019},and the Solar Wind Electrons Alphas and Protons \citep[SWEAP:][]{Kasper2016}.FIELDS is composed of several magnetometers designed to measure the vector components of the magnetic field. They are oriented to have two three-axis fluxgate magnetometers oriented in a cross-like configuration $90\degree$ from each other extending outward from the spacecraft, and one three-axis search coil magnetometer mounted to the spacecraft itself as a reference. The Flux gate magnetometers can take anywhere from 2 to 293 samples per second using the spacecraft’s coordinate system as a reference, with a range from the scale of \textit{nT} to $\mu T$ \citep{Bale2016}.
The \isis\ instrument suite consists of two instruments, Energetic Particle Instrument-Low energy (EPI-Lo) to handle the lower energy particles and EPI-Hi that detects the high energy particles \citep{McComas2016,McComas2019}, measuring a wide variety of energetic particle fluxes and pitch angles (the angle of a particle relative to the magnetic field). The EPI-Lo has an octagonal dome body supporting 80 viewfinders each supporting an 11.25\degree\ field of view in the azimuthal direction and 32\degree\ in the polar direction, to support a full 360\degree\ field of view in the azimuthal plane and a 10\degree\ to 170\degree\ field of view in the polar plane. The EPI-Hi has three telescopes, a double-ended high energy telescope (HET), a double-ended low energy telescope (LET1), and a single-ended low energy telescope (LET2) which are mounted together in a box such as to present a field of view made up of a series of five overlapping 45\degree\ half-angle cones. Three of these cones come from LET1 and LET2, with the remaining two coming from HET. Together these provide full energy coverage in the sunward and anti-sunward directions. EPI-Lo is capable of measuring ions from 20 keV/nucleon up to 15 MeV total energy and EPI-Hi is capable of measuring ions from 1-200 MeV/nucleon. These instruments can detect up to 100,000 particles per second \citep{McComas2016, McComas2019}).

The SWEAP \citep{Kasper2016} consists of the Solar Probe Cup (SPC) and the Solar Probe Analyzers (SPAN-A and SPAN-B). The Solar Probe Analyzers are electrostatic analyzers and SPC is a Faraday cup. The SPAN-A is designed to study both ion and electron spectra and can make one spectrum out of 32 survey spectra every 15 minutes and has an energy range of 100~eV to 30~keV. The SPAN-B is designed to study only electron spectra and can also make one spectrum out of 32 survey spectra every 15 minutes and has an energy range of 1~eV to 5~keV. Both these instruments have an angular range of +/-60\degree\ in elevation and 247.5\degree\ in azimuth with a partial obstruction in the first 8\degree\ of the azimuthal direction due to the thermal protection shield. The azimuth is divided into sixteen 15.5\degree\ wide angular bins and the elevation into eight 15\degree\ wide angular bins \citep{Kasper2016}. The SPC is designed to study solar wind parameters such as temperature, density, and velocity concerning both electrons and ions in the solar wind. It can make 17,580 measurements per minute and has an energy range of 100~eV to 6~keV. Since it is a Faraday cup, it does not use angular binning \citep{Case2020}. The SPC is also capable of making measurements of current, by seeing how many particles hit a collector plate over a certain period, measured by a voltage across a resistor directly attached to the collector plate. This voltage is directly proportional to the current via Ohm’s law, and the current is written in units of volts in SPC data files as a sort of “shorthand” \citep{Case2020}.

\subsection{Method Adopted} \label{sub:method}

In this study we aimed to accomplish the following objectives: 
\begin{inparaenum}
  \item To demonstrate that a subset of magnetic switchbacks are generated via interchange reconnection, and
  \item To determine whether the switchbacks within this subclass are more in line with the \textit{linear} or the \textit{flux rope} models for magnetic switchbacks.
\end{inparaenum}
To accomplish the first goal, we identified events such as CMEs, solar flares, or high speed streams in the solar corona that produce suprathermal alpha particles associated with magnetic reconnection and possess a magnetic field orientation. This orientation is outlined in Figure \ref{fig:Model} panel [a]. Given that this orientation contains open and closed field lines in close proximity, this can lead to interchange reconnection.

To identify the association between switchbacks with suprathermal alpha particles and solar activity such as CMEs, solar flares, and high-speed streams, we identified where concentrations of suprathermal alpha particles exist. Specifically, we identified spikes in alpha particle flux above 20~keV/nucleon or $\sim 84.45$~keV for an alpha particle. This energy level is indicative of ions accelerated near the Sun via reconnection events in the corona such as impulsive flare events, high-speed streams, and CMEs\citep{Reames1997,Kahler2019, Desai2020}. Given that switchbacks typically possess a proton energy range of 500~eV/particle to 2~keV/particle \citep{Mozer2020} where protons dominate the ion species in the switchback, particles above 20~keV/nucleon represent a large increase from the typical energy range of switchbacks. Switchbacks which contain elevated alpha particles in this energy range represent a particularly energetic subclass of switchbacks that display distinct physics, hereby dubbed by the authors “alpha-heavy switchbacks”. This energy range also corresponds with EPI-Lo, which, as mentioned in the previous section, measures particles above 20~keV/nucleon. 

To identify switchbacks with detectable alpha particles above 20~keV/nucleon but below 15~MeV/nucleon, we used EPI-Lo alpha particle channel Level~2 flux data from August 2018 to August 2022. Despite the ubiquity of alpha particles in the solar wind in general, they are relatively rare above 20~keV/nucleon. Alpha particles at these energies tend to occur in isolated "spikes" where the alpha particle levels detected are substantially higher for short intervals of time. We searched for these ``alpha spikes” in the the EPI-Lo data \citep{McComas2016,McComas2019} and compiled a list of all the alpha spikes within this time frame. 

\begin{figure}[htbp]
  \centering
  \centerline{
  \includegraphics[width=0.9\textwidth,clip=]{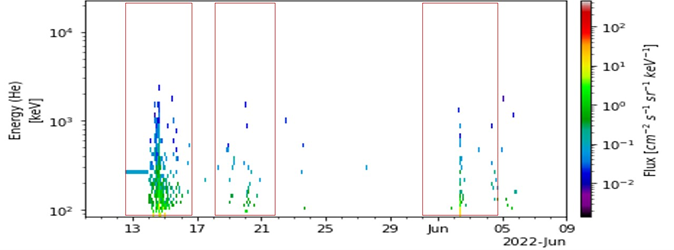}}
  \caption{ $PSP$ \isis\ EPI-Lo  Level~2 (L2) Ion Composition (IC) alpha flux from May 10 to June 10, 2022. The range of which alpha-heavy switchbacks may be found around each alpha spike is marked in red.} 
          
          \label{fig:Alpha}
\end{figure}

\begin{figure}[htbp]
  \centering
  \centerline{
  \includegraphics[width=1.0\textwidth,clip=]{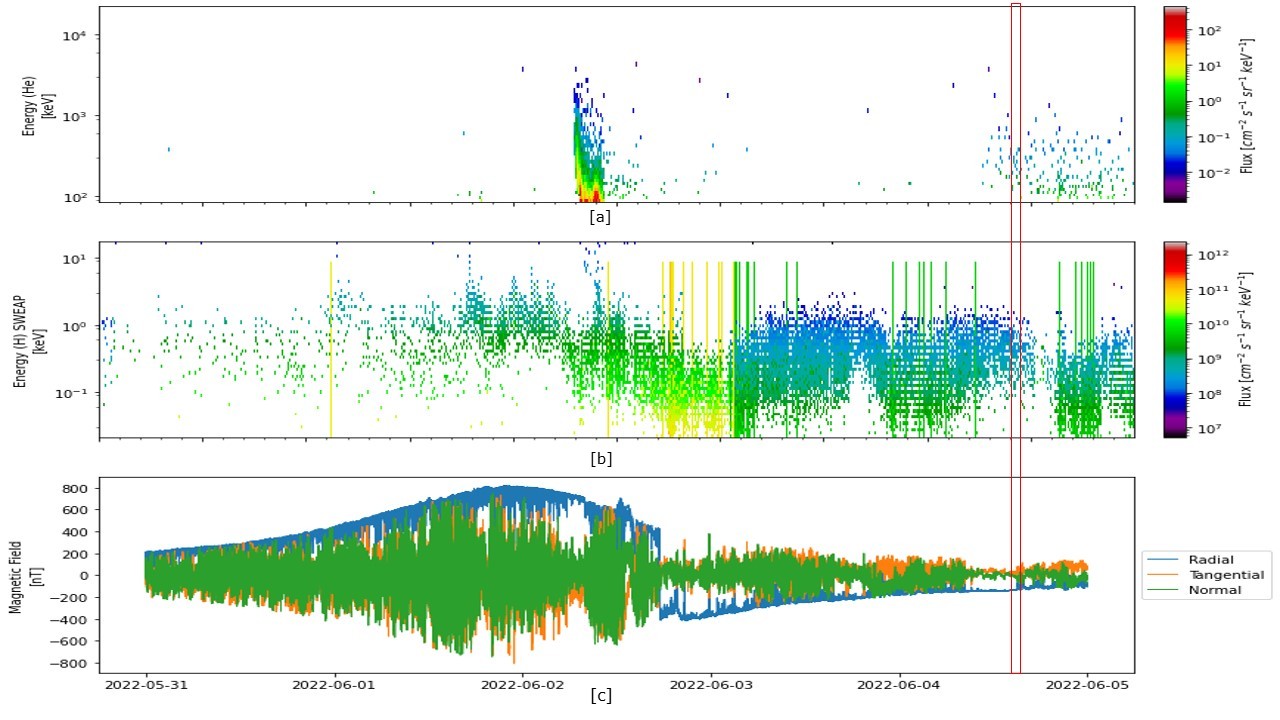}}
  \caption{ Parker Solar Probe data for an alpha spike over the period of May~31 to June~5, 2022. Panel~(a): The \isis/EPI-Lo Level~2 IC alpha flux. Panel~(b): SWEAP Thermal Proton Flux. Panel~(c): The FIELDS magnetic field data. The location of an alpha-heavy switchback may be found marked in red.
          } 
          \label{fig:Alpha_Example}
\end{figure}

Alpha spikes themselves tend to occur at very infrequent intervals, with only a few appearing per month during the most active periods of the data. They also may be of varying duration and may be either concentrated over a few hours or may be more diffuse lasting for several days. Because of this, we searched for alpha heavy switchbacks within a 48-hour window before and after the point of highest alpha flux in the alpha spike (hereafter referred to as the alpha spike peak) using Level~2 data in a radial (R), tangential (T), normal (N) (RTN) coordinate system from the Flux Gate Magnetometer \citep{Jannet2020} of the FIELDS instrument suite \citep{Bale2016}. Figure~\ref{fig:Alpha} depicts the typical scarcity of alpha spikes, as well as the range in which we searched for alpha-heavy switchbacks surrounding the alpha spikes. This range is marked in red.These alpha spikes also show the physical signatures of being part of a larger structure. Figure \ref{fig:Alpha_Example} depicts an example of an alpha spike (Panel~(a)) with the magnetic signature of a larger structure (Panel~(c)) and the corresponding thermal ion population around the alpha spike (Panel~(b)). An associated alpha-heavy switchback to this alpha spike is marked in red, however it is on a relatively small scale compared to the magnetic field surrounding the alpha spike and is thus not easily seen on this scale. Therefore, We attempted to identify these alpha spikes with known physical processes. We accomplished this by looking for transient coronal activity such as CMEs, solar flares, or high-speed streams occurring at the temporal and spatial vicinity of the spikes recorded at $PSP$ which we defined as an alpha spike being detected at $PSP$ within 2~hours of a CME or high-speed stream being known to have been detected at $PSP$. 

We identified CMEs, solar flares, and high-speed streams at $PSP$ in the NASA Goddard Database of Notifications, Knowledge, Information (DONKI) database. In the case of a solar flare, energetic ions may accelerate at the solar surface and propagate through the solar wind, only arriving at the detector long after the flare is initially observed via a burst of electromagnetic radiation. In such a case, the ions will be at a variety of energy levels and will thus be moving at different speeds in the solar wind, arriving at a probe like $PSP$ at different points. With that in mind, an alpha spike may be associated with a solar flare if 50\% or more of the suprathermal alpha particles making up the alpha spike are at an energy level where the calculated time for alpha particles at that energy to travel to $PSP$ from a known solar flare on the Sun is within 2~hours of the peak of the observed alpha spike. In the case of high-speed streams, they (high-speed streams) may be predicted somewhat based on the observation of coronal holes, but DONKI databases are typically in-situ observations by probes like $PSP$ \citep{David2022}. In the case of CMEs, those in the DONKI database are projections based on ENLIL-WSA+CONE modelling, which is a model where solar wind conditions are estimated via observations of the local solar magnetic field and these conditions are projected outward as they move through space \citep{Mays2015}. Because CMEs in the DONKI database are projections rather than documented events recorded at $PSP$, we cross-referenced the DONKI predictions with known CMEs at $PSP$ outlined in \cite{Salman2024} and the Large-Scale Structures Originating from the Sun  (LASSOS) ICME catalog to ensure the alpha spikes are being compared to a real event. We further confirmed the CME was observed by $PSP$ by examining the thermal alpha population for alpha particles around 1~keV surrounding the alpha spike because thermal alpha particles around 1~keV in the solar wind are typical of CMEs \citep{Reinard2001}. Finally, we tested the plasma data for shock conditions in the Rankine-Hugonoit equations \begin{equation} 
\left[ \rho v_n \right] = 0
\label{eq:3}
\end{equation} 
\begin{equation} 
\left[ \rho v_n^2 + p + \frac{B_t^2}{2\mu_0} \right] = 0 
\label{eq:4}
\end{equation} 
\begin{equation} 
\left[ \rho v_n v_t - \frac{B_n B_t}{\mu_0} \right] = 0 
\label{eq:5}
\end{equation} 
\begin{equation} \left[ v_n \left( \frac{1}{2} \rho v^2 + \frac{\gamma}{\gamma - 1} \frac{p}{\rho} + \frac{B^2}{2\mu_0} \right) \right] = 0 
\label{eq:6}
\end{equation}
\begin{equation} 
\left[ B_n \right] = 0
\label{eq:7}
\end{equation} 
\begin{equation} 
\left[ v_n B_t - v_t B_n \right] = 0 
\label{eq:8}
\end{equation} where applicable to determine the shocks associated with a CME or high-speed stream. In these equations, $\rho$ is the density, $v_{n}$ is the velocity component of the plasma perpendicular to the shock, $v_{t}$ is the velocity component of the plasma tangential to the shock, $p$ is the plasma pressure, $\mu_{0}$ is the vacuum permittivity, $B_{n}$ is the magnetic field component of the plasma perpendicular to the shock, $B_{t}$ is the magnetic field component of the plasma tangential to the shock, and $\gamma$ is 5/3.

Although the definition of switchbacks is not precisely agreed upon, for the purpose of identifying switchbacks in our studies, we adopted the following criteria. The R-component and at least one other component of the magnetic field in the RTN coordinate system changed polarity for a duration of at least 5~minutes, but not more than 3.5 hours. This duration was chosen to ensure stability of the switchbacks, and to more reliably ensure correlation with other types of solar activity compared to shorter duration switchbacks \citep{Horbury2023}. Further, the polarity reversal must meet the criterion such that the magnetic field components that comprise the event switch direction such that the normalized deflection measure \( z > 0.5\) where the equation for \( z \) is 
\begin{equation}
    z = \frac{1}{2} \cos (\alpha) >0.5,    \label{eq:9}
\end{equation}
where \( \alpha\) is the angle between the pointwise magnetic field and a local average in at least two coordinates, following the definition used in \cite{DudockdeWit2020} and  \cite{Pecora2022}. In order to conclusively show that these alpha spikes and alpha-heavy switchbacks are more directly correlated compared to the general switchback population, We drew a Chi-square analysis between the switchbacks and the alpha spikes such that we calculate the parameter 
\begin{equation} 
\chi^2 = \sum_{i} \frac{(O_i - E_i)^2}{E_i} 
\label{eq:10}
\end{equation}
where \(O\) is the observed frequency of an event and \(E\) is the expected frequency of observing an event. The expected frequencies were calculated using the equation
\begin{equation}
E_{(type, spike\_status)} = \frac{\text{Total Switchbacks of Type} \times \text{Total Switchbacks with Spike Status}}{\text{Total Switchbacks}}
\label{eq:11}
\end{equation}
. These expected frequencies were calculated  for regular switchbacks associated with a spike, regular switchbacks without a spike, alpha-heavy switchbacks with a spike, and alpha-heavy switchbacks without a spike. Each of these were calculated over two different time periods. The first consisted of a 3~month period of low alpha spike activity from October 2018 to December 2018, and the second consisted of a 3~month period of high alpha spike activity from June 2022 through August 2022. 

In a Chi-square analysis, the degrees of freedom are specified via the equation: \begin{equation} df = (r - 1) \times (c - 1) \label{12} \end{equation}

where r is the number of rows in the data table, and c is the number of columns in the data table. The value of df helps determine whether there is any association between variables. For example, a df of 1 indicates that there is only one independent comparison being made between the observed and expected frequencies, making it straightforward to evaluate the relationship between two variables. We calculate this parameter over both 3~month periods to provide clarity that this is a simple test of association between variables and to provide context for interpreting the results.

To accomplish the second objective, we sought to determine whether these alpha-heavy switchbacks display traits more in-line with fast magnetosonic waves, with flux ropes, or some other model. Once we identified a catalog of alpha-heavy switchbacks and associated alpha spikes, we checked if these switchbacks display properties of fast magnetosonic waves using the following method: because the speed of fast magnetosonic waves can be calculated using equation \ref{eq:2}, if we use typical values for the solar corona where the Alfv\'{e}n velocity is approximately
\begin{equation}
    v_A = 1000 \, \kms
    \label{eq:13}
\end{equation}
and the speed of sound is:
\begin{equation}
    c_s = 200 \, \kms
    \label{eq:14}
\end{equation}
This would yield a speed for the fast magnetosonic wave of 
\begin{equation}
    c_ms=1019 \, \kms
    \label{eq:15}
\end{equation}
Therefore, the arrival time for a fast magnetosonic wave would have an error of
\begin{equation} \Delta t \approx \left| \frac{d}{c _{m s}^2} \right| \Delta v + \frac{1}{c_{m s}} \Delta d + \Delta t_{\text{inst}}
\label{eq:16}
\end{equation}
Where  \( d \) is the distance to the spacecraft, \( \Delta d \) is the error in the distance, \( \Delta v \) is the error in the velocity, and \( \Delta t_{\text{inst}} \) is the instrumentation timing precision. 

If the arrival time of a magnetic switchback at $PSP$ from the Sun is significantly outside the error in equation \ref{eq:15}, it is unlikely that the switchbacks are fast magnetosonic waves. From there, we applied the W\`{a}len test to test the presence of Alfv\'{e}nic flows as might be predicted from Alfv\'{e}nic turbulence models by using the relation 
\begin{equation}
V = \frac{1}{2} V_A
\label{eq:17}
\end{equation}
where $V$ is the plasma inflow velocity to the current sheet and $V_{A}$ is the Alfv\'{e}n velocity. If this relation holds, it suggests that the plasma fluctuations are Alfv\'{e}nic and may indicate Alfv\'{e}nic turbulence \citep{Hudson1970}.

To calculate the inflow velocity, we used the relation 
\begin{equation}
V_{\text{in}} = \frac{\Delta B}{B_{\text{up}}} \cdot V_A
\label{eq:18}
\end{equation}
where  \( B_{\text{up}} \) is the magnetic field upstream of the switchback and \( \Delta B \) is the change in the total magnetic field during the switchback.

We also analyzed the solar wind properties of the switchback plasma. Specifically, we used SPC Level~3 data to identify changes in the density of the plasma, temperature, and solar wind velocity in-order to determine underlying plasma structures in tandem with the magnetic field. In instances where SPC Level~3 data was unavailable due to quality flags, we used SWEAP SPAN-AI data instead. From there, we used the pitch angle data from the Level~2 EPI-Lo instrument to develop a pitch-angle distribution to analyze the particle dynamics within the structure of the switchbacks. Finally, we analyzed the behavior of the radial magnetic field and the radial solar wind velocity fluctuations during the switchbacks to look for correlations between them to determine if they display behaviors of magnetosonic waves. In so doing, we demonstrated  that the plasma dynamics within switchbacks either demonstrate a reconnection-based process largely dominated by fast magnetosonic waves, or by some other process. For determining the pitch angle distributions and thermal population during the selected switchbacks, we used the SWEAP/SPAN-A Level~2 ion data; this is because out of the two electrostatic analyzers SPAN-A and SPAN-B, only SPAN-A is equipped to handle Ions \citep{Kasper2016}.

%
%
\section{Results} \label{sec:results}
In Tables \ref{tab:table1} and \ref{tab:table1B}, we see each alpha spike ($\alpha$ spike) laid out in chronological order, along with whether they are associated with any particular type of solar activity (Associated with Solar Activity?), whether they contain an alpha-heavy switchback within a 48 hour period before and after the peak of the alpha spike (Associated $\alpha$-H SB?), the distance from the sun $PSP$ was at the moment of measurement (Dist. from Sun), and the time it would take alpha particles within EPI-Lo energy range to reach $PSP$ at this distance from the sun ($\Delta$t for $\alpha$s from sun). 

In Tables \ref{tab:table2} and \ref{tab:table2B} we see each alpha-heavy switchback indexed in chronological order, the change in temperature ($\Delta$ T) within the switchback relative to the surrounding solar wind, the change in radial velocity ($\Delta$ Vr) within the switchback relative to the surrounding solar wind, the change in velocity normal to the ecliptic plane ($\Delta$ Vn) within the switchback relative to the surrounding solar wind, the change in velocity tangential to the ecliptic plane ($\Delta$ Vt) within the switchback relative to the surrounding solar wind, the change in proton density ($\Delta$ n) within the switchback relative to the surrounding solar wind, the duration of the switchback per its magnetic field (SB dur), the duration of the changes in the radial velocity ($\Delta$ V dur), the time difference between the alpha-heavy switchback and the alpha spike where + is after the alpha spike and - is prior ($\Delta$ t $\alpha$ spike), whether the alpha-heavy switchback displays behavior indicative of field-aligned currents (FA Currents?), and whether or not the alpha-heavy switchback displays behavior indicative of magnetic mirroring (Mirrors?). Here, unavailable data is marked as N/A. 

The alpha spikes from 2018 through 2019 are very rare with only four events occurring prior to 2020 but increasing in frequency from 2020 onward. Some alpha spikes, including those identifiable with solar activity, displayed more than one alpha-heavy switchback separated by more than 3.5 hours, and several others displayed multiple alpha-heavy switchbacks near each other, often occurring in “chains” where one immediately follows another. Because of this, we separated the instances of switchbacks separated by more than 3.5 hours that occur on the same day in Tables \ref{tab:table2} and \ref{tab:table2B} via a number, so that two alpha-heavy switchbacks occurring on a single day occurring 6 hours apart might be recorded as MM-DD-YYYY (1) and MM-DD-YYYY (2). “Chains” of switchbacks were grouped together and considered a single switchback for the purposes of Tables \ref{tab:table2} and \ref{tab:table2B}. 

\begin{figure}[htbp]
  \centering
  \centerline{
  \includegraphics[width=0.95\textwidth,clip=]{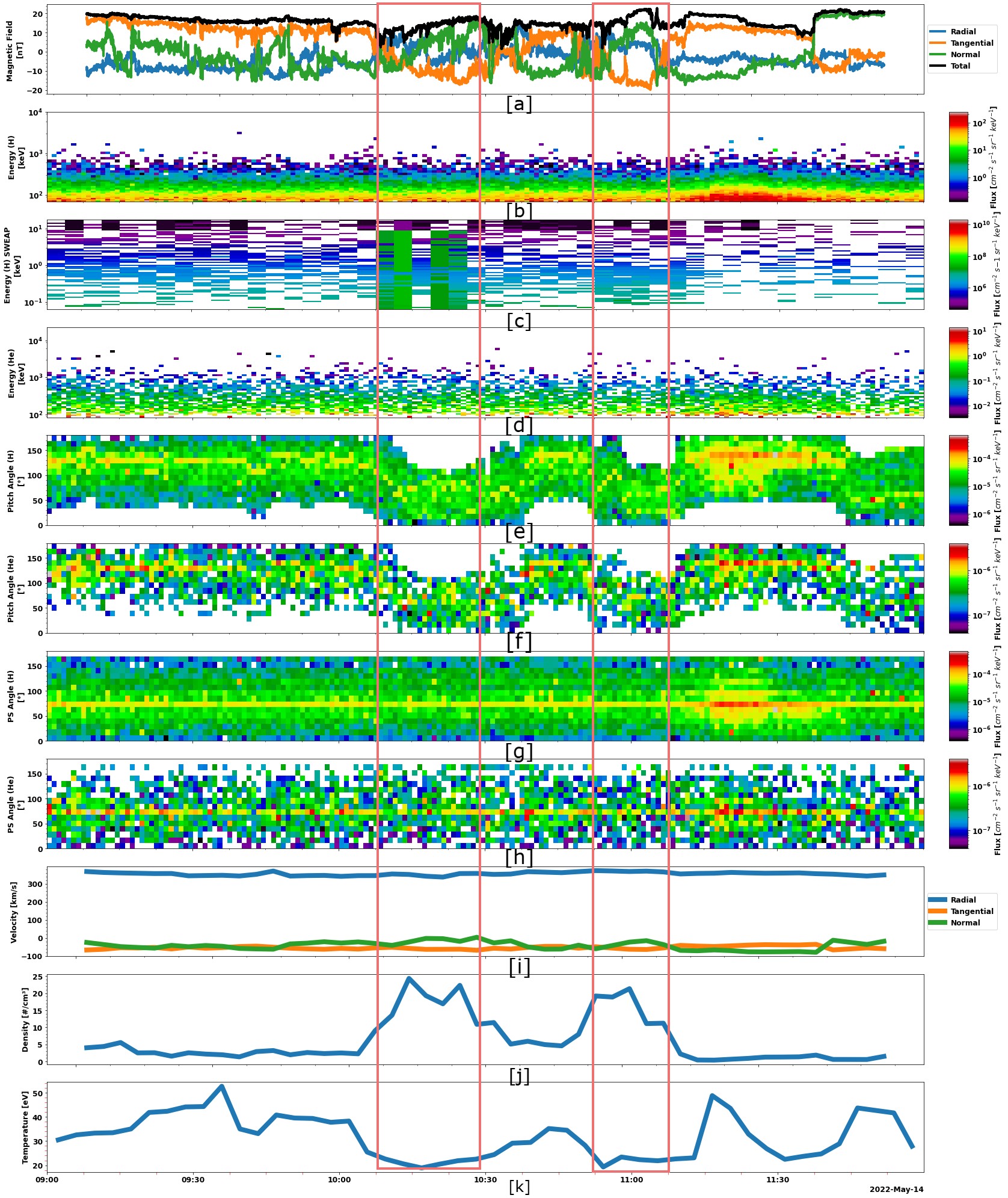}}
  \caption{ Plots of a switchback. Panel~(a): FIELDS magnetic field data in RTN coordinates. Panel~(b): EPI-Lo proton energy-flux spectrogram. Panel~(c): SWEAP proton energy-flux spectrogram. Panel~(d): EPI-Lo alpha energy-flux spectrogram. Panel~(e): proton pitch angle distribution. Panel~(f): Alpha pitch angle distribution. Panel~(g): proton Parker spiral angle distribution. Panel~(h): Alpha Parker spiral angle distribution. Panel~(i): Solar wind velocity. Panel~(j): Proton density. Panel~(k): ion temperature.
          } 
          \label{fig:Switchback_Spectrogram}
\end{figure}

In Figure \ref{fig:Switchback_Spectrogram} we see the proton and alpha flux in a typical alpha-heavy switchback associated with an alpha spike. Panel~(a) depicts the boundaries of the switchback determined by the magnetic field, Panel~(b) depicts an energy-flux spectrogram for suprathermal protons, Panel~(c) depicts an energy-flux spectrogram for thermal protons, Panel~(d) depicts an energy-flux spectrogram of suprathermal alpha particles, Panel~(e) depicts a pitch angle distribution for suprathermal protons, Panel~(f) depicts a pitch angle distribution for suprathermal alpha particles, Panel~(g) depicts a Parker Spiral angle (angle of the flow of ions relative to the Parker Spiral) distribution for protons, Panel~(h) depicts a Parker Spiral angle distribution for alpha particles, Panel~(i) depicts the solar wind velocity in RTN coordinates, Panel~(j) depicts the solar wind proton density, and Panel~(k) depicts the solar wind ion temperature.

Table \ref{tab:table3} depicts six months of data, the first three (10-12 2018) depict a time of low solar activity, and the last three (06-08 2022) depict a higher period of solar activity. This table also shows the corresponding number of alpha spikes (\# of $\alpha$ Spikes), the number of regular switchbacks (\# of Regular Switchbacks), and the number of alpha-heavy switchbacks (\# of $\alpha$-H SBs) during each of these months.

The alpha-heavy switchbacks in Table \ref{tab:table3} are statistically associated via the Chi-square analysis 
\begin{equation} 
E_{\text{reg, spike}} = \frac{165 \times 2}{169} \approx 1.95 
\label{eq:19}
\end{equation} 
\begin{equation} 
E_{\text{reg, no spike}} = \frac{165 \times 167}{169} \approx 163.05 
\label{eq:20}
\end{equation} 
\begin{equation} E_{\text{alpha-heavy, spike}} = \frac{4 \times 2}{169} \approx 0.05 
\label{eq:21}
\end{equation} 
\begin{equation} E_{\text{alpha-heavy, no spike}} = \frac{4 \times 167}{169} \approx 3.95 
\label{eq:22}
\end{equation} 

\begin{equation} 
\chi^2_{\text{reg, spike}} = \frac{(1 - 1.95)^2}{1.95} = \frac{0.9025}{1.95} \approx 0.46 
\label{eq:23}
\end{equation} 
\begin{equation} 
\chi^2_{\text{reg, no spike}} = \frac{(164 - 163.05)^2}{163.05} = \frac{0.9025}{163.05} \approx 0.01
\label{eq:24}
\end{equation} 
\begin{equation}
\chi^2_{\text{alpha-heavy, spike}} = \frac{(1 - 0.05)^2}{0.05} = \frac{0.9025}{0.05} = 18.05 
\label{eq:25}
\end{equation}
\begin{equation} 
\chi^2_{\text{alpha-heavy, no spike}} = \frac{(3 - 3.95)^2}{3.95} = \frac{0.9025}{3.95} \approx 0.23 
\label{eq:26}
\end{equation} 
\begin{equation} 
\chi^2_{\text{total}} = 0.46 + 0.01 + 18.05 + 0.23 = 18.75 
\label{eq:27}
\end{equation} 
\begin{equation} 
df = (2 - 1) \times (2 - 1) = 1 
\label{eq:28}
\end{equation}
for the 3 month period in 2018 and the Chi-square analysis 
\begin{equation} 
E_{\text{reg, spike}} = \frac{196 \times 50}{246} \approx 39.84 
\label{eq:29}
\end{equation} 
\begin{equation}
E_{\text{reg, no spike}} = \frac{196 \times 186}{246} \approx 147.92 
\label{eq:30}
\end{equation} 
\begin{equation}
E_{\text{alpha-heavy, spike}} = \frac{40 \times 50}{246} \approx 8.13 
\label{eq:31}
\end{equation} 
\begin{equation}
E_{\text{alpha-heavy, no spike}} = \frac{40 \times 186}{246} \approx 30.24 
\label{eq:32}
\end{equation} 
\begin{equation} 
\chi^2_{\text{reg, spike}} = \frac{(10 - 39.84)^2}{39.84} \approx 24.34 
\label{eq:33}
\end{equation} 
\begin{equation} 
\chi^2_{\text{reg, no spike}} = \frac{(186 - 147.92)^2}{147.92} \approx 9.79
\label{eq:34}
\end{equation} 
\begin{equation} 
\chi^2_{\text{alpha-heavy, spike}} = \frac{(40 - 8.13)^2}{8.13} \approx 121.70 
\label{eq:35}
\end{equation} 
\begin{equation} 
\chi^2_{\text{alpha-heavy, no spike}} = \frac{(0 - 30.24)^2}{30.24} \approx 30.24 
\label{eq:36}
\end{equation} 
\begin{equation} 
\chi^2_{\text{total}} = 24.34 + 9.79 + 121.70 + 30.24 = 186.07 
\label{eq:37}
\end{equation} 
\begin{equation} 
df = (2 - 1) \times (2 -1)=1
\label{38}
\end{equation}
for the 3 month period in 2022.

Finally, Tables \ref{tab:table4} and \ref{tab:table4b} depict each of the alpha-heavy switchbacks, whether they possess a fluctuation in the radial component of the magnetic field (Br fluct.), whether these magnetic field fluctuations correlate (Yes), anticorrelate (Yes (-)), or both (Yes(+/-)) with the radial component of the solar wind velocity (Corr. with Vr), the distance from the sun $PSP$ was at the moment of measurement (Dist. from Sun), and the time it would take alpha particles within EPI-Lo energy range to reach $PSP$ at this distance from the Sun ($\Delta$t for $\alpha$s from sun), whether these switchbacks pass the W\'{a}len test (Pass W\'{a}len Test), and whether the bulk of the alpha particles have transit times from the sun that are outside the error for a fast magnetosonic wave ($>$ err for FMSW).

\section{Discussion} \label{sec:discussion}

Of the ``alpha spikes" that represent elevated levels of suprathermal alpha particles, 84.1\% of the alpha spikes were detectable with a specific CME, solar flare, or high-speed stream. Of the alpha spikes associated with specific solar processes, 24.5\% were associated with CMEs, 60.4\% were associated with solar flares, and 15.1\% were associated with high-speed streams. Alpha spikes associated with CMEs were more likely to be associated with an alpha-heavy switchback than alpha spikes associated with solar flares. 69.2\% of the CME associated alpha spikes were associated with an alpha-heavy switchback, and 69\% of the solar flare associated alpha spikes were associated with an alpha-heavy switchback. Only 37.5\% of the alpha spikes associated with high-speed streams were associated with an alpha-heavy switchback. Each one of these ``alpha spikes" is observable from 0.10~AU to 0.86~AU. These statistics are reflected in Table \ref{tab:table1}. 

Alpha-heavy switchbacks in most cases occurred within one day or a few hours before or after the alpha spike, with the average gap between an alpha spike and alpha-heavy switchback being 22.62 hours ($\Delta t$ alpha spike). Although not every alpha spike had an associated alpha-heavy switchback, in several occasions the reason we were unable to find an alpha-heavy switchback was the existence of data gaps in the \isis$\;$or FIELDS measurements. Therefore, it is possible that an alpha-heavy switchback did occur but did not get recorded during these time frames because of the data gaps.

The alpha-heavy switchbacks we observed were anywhere from 10 minutes to 3.5 hours in duration with an average duration of 60.43 minutes. All of them displayed perturbation in the solar wind parameters such as velocity, temperature, or density that were larger or smaller than the surrounding solar wind, but 29.6\% of them displayed such perturbations at intervals that did not directly correspond with the duration of the switchback itself. 87.5 \% of these cases where the perturbations did not directly correspond with the duration of the switchback itself involved the plasma perturbation being longer than the switchback duration, as was the case in Figure \ref{fig:Switchback} (Panels~(a) and (b)). In such switchbacks, the average perturbed plasma velocity duration was 35 minutes longer than the associated alpha-heavy switchback. Additionally, while in most alpha-heavy switchbacks the temperature of the switchback was in fact significantly higher than the surrounding plasma, in these cases of the perturbed plasma velocity duration extending past the magnetic switchback boundary, the temperature inside the switchback was lower than the surrounding plasma. These trends are laid out in Tables \ref{tab:table2} and \ref{tab:table2B}. 

The statistics presented in Tables \ref{tab:table1} and \ref{tab:table1B} indicate that elevated levels of suprathermal alpha particle flux were primarily associated with solar events like CMEs, solar flares, and high-speed streams, and suggest these ions were not there coincidentally, but were there for a reason connected to the magnetic structure of the Sun. Solar events associated with alpha-heavy switchbacks were relatively few in 2018, 2019, and 2020 but more frequent in 2021, 2022, 2023, and 2024. This could be explained by the period around 2020 corresponding with the solar cycle exiting the solar minimum and heading towards the solar maximum. As the Sun got closer to solar maximum, there were clearer low-latitude coronal holes causing a peak in stream interaction regions and interchange reconnection at the boundaries \citep{Sanchez-Garcia2023}.

Further, although alpha-heavy switchbacks did not exclusively occur within 48 hours before and after a switchback, they also tended to cluster around the alpha spikes within this time frame, and seemed to increase in frequency in correlation with an increase in solar activity moving from 0-0.5 events per week to 2.25-5.25 events per week. Meanwhile, non-alpha-heavy switchbacks occurred at a rate of approximately 15-16 events per week regardless of solar activity as depicted in Table \ref{tab:table3}.

Given this data, the Chi-squared analysis showed an association between the alpha spikes and the alpha-heavy switchbacks that positively correlated with increased solar activity, while non-alpha-heavy switchbacks showed no significant correlation with increased solar activity. This suggests that the magnetic reconnection processes that produce solar flares, CMEs, and high-speed streams may produce alpha-heavy switchbacks, but that alpha-heavy switchbacks represent a separate population from the rest of the switchbacks. Whether alpha-heavy switchbacks represent different physics however, is a matter for a future study. 
 
The increase in solar activity during this period would cause more mixing of fast and slow solar wind streams. While this fact in isolation wouldn’t contradict theories of Alfv\'{e}nic turbulence or shear forces between the fast and slow solar wind driving the switchback phenomenon \citep{McComas2003}, the fact that the switchbacks did not pass the W\`{a}len test indicates that they are not sufficiently driven by Alfv\'{e}nic flow as to be explainable as a form of Alfv\'{e}nic turbulence. Moreover, the observed fluctuations in the radial magnetic field direction seen in our study of alpha-heavy switchbacks is not an Alfv\'{e}nic property and is characteristic of magnetosonic waves \citep{Lighthill1960}. 
Additionally, given that EPI-Lo has a single count limit of at least $10^{-4} cm^{-2}s^{-1}sr^{-1}keV^{-1}$ \citep{McComas2016}, these alpha particle fluxes within switchbacks were relatively low density in the switchback plasma, but still clearly showed a lack of mirroring in the pitch angle distributions. This reinforces work done by \cite{Bandyo2021}, which noted that the gyroradii of suprathermal ions in magnetic switchbacks are either similar or of much larger size than the radius of curvature of magnetic switchbacks. This is much easier to reconcile with the \cite{Zank2020} model, than a situation where a flux rope produced by reconnection would see suprathermal ions accelerated by the same reconnection event be unable to mirror within the flux rope. Furthermore, we observed a correlation between the radial magnetic field fluctuations and the fluctuations in the radial solar wind velocity, which is also highly indicative of fast magnetosonic waves \citep{Zank2020}. This is supported by the fact that the pitch angle distributions within the switchbacks showed no evidence of mirroring or field aligned currents as we would expect given the Drake model. The switchbacks of which an example is depicted in Figure \ref{fig:Switchback_Spectrogram} showed a relatively uniform distribution across the switchbacks when plotting ion flux relative to the Parker angle. This is much easier to reconcile with the \cite{Zank2020} model than the \cite{Drake2021} model, due to the fact that fast magnetosonic waves generally propagate with the Parker Spiral due to their phase speed being faster than the Alfv\'{e}n speed \citep{Ofman2023}, but a flux rope as predicted in \cite{Drake2021} would have a more twisted magnetic field orientation, causing significant differences in angle relative to the Parker Spiral. The changes in solar wind density, temperature, and velocity that we see are largely confined within the magnetic boundaries of the switchbacks, and are consistent with known behavior of fast magnetosonic waves to produce regions of both compression and rarefaction \citep{Ofman2023}. 
 
There are some cases of alpha-heavy switchbacks for which the changes in velocity do not match up with the duration of the switchback itself, which is not necessarily indicative of fast magnetosonic waves, but these may be indicative of radiative cooling, adiabatic expansion, or the interaction of the switchback with a slow magnetosonic wave or Alfv\'{e}n wave \citep{Ofman2023}. We also observed anticorrelations between the radial magnetic field fluctuations and the fluctuations in the radial solar wind velocity, which while not necessarily indicative of magnetosonic waves could be indicative of alternative modes such as Alfv\'{e}n waves, or it could indicate complex interactions in the plasma, such as those caused by inhomogeneities or standing wave patterns \citep{Archer2023}. While the fact that the alpha-heavy switchbacks did not pass the W\`{a}len test makes it hard to reconcile the data with a model including Alf\'{e}nic waves interacting with fast magnetosonic waves, this does not necessarily preclude their presence. The anticorrelation between the radial magnetic field and solar wind velocity components is also consistent with cases of phase mixing between fast magnetosonic waves and Alfv\'{e}n waves \citep{Nakariakov1997}. Such a case of phase mixing would not be likely to pass the W\`{a}len test due to the flow not being purely Alfv\'{e}nic, but could help to explain why many switchbacks have been observed to display some Alfv\'{e}nic properties. 

Only one alpha-heavy switchback showed no correlation between the radial magnetic field fluctuations and the radial solar wind velocity fluctuations at all. The alpha heavy switchbacks that displayed an anticorrelation between the radial magnetic field fluctuations and the fluctuations in the radial solar wind velocity were on average seen at 0.596 AU from the Sun, while those that displayed a correlation showed an average distance of 0.598 AU from the Sun. Therefore, because all alpha-heavy switchbacks were observed between 0.13 AU and 0.81 AU, it can be inferred that if the anticorrelated alpha heavy switchbacks do represent phase mixing between fast magnetosonic waves and Alfv\'{e}n waves, they show no substantial difference in how far they may propagate relative to alpha-heavy switchbacks propagating as a pure fast magnetosonic wave up to 0.81 AU. 

Finally, as depicted in Tables \ref{tab:table4} and \ref{tab:table4b}, each of the switchbacks possess a property where they show fluctuations in the radial magnetic field components that correlate with the radial velocity vector. Additionally, 56\% of the alpha-heavy switchbacks for which velocity data was available showed a correlation between the radial magnetic field fluctuations and the fluctuations in the radial solar wind velocity. However, 39\% showed an anticorrelation between the radial magnetic field fluctuations and the fluctuations in the radial solar wind velocity.

\section{Conclusion} \label{sec:conclusion}

Magnetic Switchbacks are quick reversals of the heliospheric magnetic field directions, also known as the ``s-shaped kinks" detected first in the Ulysses data and later in other spacecraft data such as Parker Solar Probe, Solar Orbiter, and Helios \citep[e.g.][and the references therein]{Balogh1999,Goldstein1995,Borovsky2016,Horbury2018,Mariani1979,Bale2019,Kasper2019,Bowen2020, McComas2019,Schwadron2021,DudockdeWit2020,Horbury2020,Mozer2020,Rouillard2020,Tenerani2020}. Several models and theories have been proposed to explain this observed phenomenon \cite{Drake2021, Zank2020} and our aim in the present analysis was to investigate the link between switchbacks and interchange reconnection. In particular, we sought to statistically tie some switchbacks to magnetic reconnection processes and to observe evidence of field-aligned currents or lack thereof in the pitch angle distributions of these switchbacks so as to provide evidence for either the \citet{Zank2020} model or \citet{Drake2021} model. In so doing, we aim to provide further evidence for magnetic interchange models of switchbacks that predict either the absence or presence of field-aligned currents. In this effort, we compiled all available $PSP$ data that has sufficient  suprathermal alpha flux  within the switchbacks that allow us to tie switchbacks to  magnetic reconnection processes that can produce suprathermal alpha particles.

We summarize our main findings below:
\begin{enumerate}
  \item Switchbacks containing suprathermal alpha flux correlated with increased solar activity such as CMEs, 
        high-speed streams, and solar flares. At the same time, switchbacks not meeting this criteria appear 
        to occur at rates that are unaffected by increasing solar activity which implies the existence of 
        multiple populations of switchbacks with differing physical mechanisms behind them. 
  \item Additionally, the fact that these  "alpha-heavy" switchbacks primarily take place in the vicinity 
        of magnetic reconnection driven solar activity such as a CME, high-speed stream, or solar flare suggests that there is a connection between magnetic reconnection and alpha-heavy switchbacks. These types of solar activity may possibly produce interchange reconnection as a result of the magnetic field orientation involving open and closed field lines
        \citep{Crooker2010, Sitnov2021} an example of which is depicted in Figure \ref{fig:Model}.       
  \item  Furthermore, the behavior of suprathermal ions within the selected switchbacks themselves 
        showed that they display a lack of evidence of field-aligned currents expected of the \citet{Drake2021} model as well as perturbations in the solar wind density, velocity, and temperature, that are more consistent with the linear model of interchange reconnection for magnetic switchbacks proposed by \citet{Zank2020} than the flux rope model proposed  by \citet{Drake2021}. 
  \item  Moreover, in a little less than half of all cases of alpha-heavy switchbacks, they display behavior 
        that could imply phase mixing of fast magnetosonic waves and Alfv\'{e}n waves, though this would require 
        future study to completely confirm. 
\end{enumerate}
However, the \citet{Zank2020} model fails to account for the instability of the magnetic kink propagating out to $PSP$ distances, which the flux rope model of \citet{Drake2021} does. Consequently, this implies that the linear interchange reconnection model is incomplete  and that there may be a yet unaccounted for interaction that allows switchbacks to propagate as a fast magnetosonic wave much further than currently predicted. Moreover, it remains unclear whether the microstreams described in the \citet{Neugebauer2021} model being produced by a flux rope model would also be consistent with the jets predicted by \citet{Zank2020}. Although the \citep{Zank2020} model of a fast magnetosonic wave generated by interchange reconnection is more consistent with  the results of our present study on alpha-heavy switchbacks, a more detailed study using additional data, if available, is required to confirm if the non alpha-heavy switchbacks are governed by the same physics. The main caveat to keep in mind is that alpha-heavy switchbacks do not encompass all switchbacks, and it is also not currently known why only certain switchbacks possess a higher alpha particle flux.

\section{Acknowledgements}

\noindent This work is supported by the National Science Foundation grant 2026579 
awarded to BP.

\noindent EM would like to give special thanks to James (Andy) Edmond for assistance with the software used in this paper. 

\noindent We would like to acknowledge David McComas at Princeton University, CDAWeb, and the FIELDS and \isis teams for providing the publicly available data for this study, as well as NASA for providing the available funding for these projects. 

 Data from $PSP$ instruments was accessed from 
\url{https://spp-isois.sr.unh.edu/data_public/ISOIS/level2/}
\url{http://sweap.cfa.harvard.edu/pub/data/sci/sweap/}
\url{http://research.ssl.berkeley.edu/data/psp/data/sci/fields/l2/mag_RTN/}.
\noindent The FIELDS, SWEAP, and \isis experiments on the Parker Solar Probe spacecraft  were designed and developed under NASA contract NNN06AA01C.

  \noindent Several plots were made using the assistance of the PYSPEDAS software. \citep{Angelopoulos2019}. 
 
 \noindent the DONKI database was accessed using the DONKI search tool \url{https://kauai.ccmc.gsfc.nasa.gov/DONKI/search/}.

 \noindent the Parker Solar Probe ICME database was accessed using the LASSOS catalog 
\url{https://science.gsfc.nasa.gov/lassos/ICME_catalogs/parker-catalog.shtml}
 

\bibliographystyle{aasjournal}
\bibliography{references.bib}
\begin{table}[htbp]
\centering
\begin{tabular}{|l|c|c|c|c|c|c|}\hline\hline
$\alpha$ spike & Associated with Solar Activity? & Associated $\alpha$-H SB? &Dist. from Sun & $\Delta$t for $\alpha$s from Sun \\ \hline\hline
        11-11-2018  & Yes (CME)& Yes & 0.24 AU &  0.17-10.6 h \\  \hline
        02-15-2019  & No  & No & 0.86 AU & 0.65-36.1 h\\  \hline
        02-18-2019  & No  & No & 0.83 AU & 0.63-34.8 h \\  \hline
        05-28-2020  & Yes (CME)& Yes & 0.34 AU & 0.25-14.2 h \\  \hline
        11-29-2020  & Yes (CME)& Yes& 0.81 AU & 0.61-33.7 h\\  \hline
        07-01-2021  & Yes (CME)& Yes& 0.75 AU & 0.55-31.3 h\\  \hline
        08-28-2021  & Yes(Solar Flare)  & Yes & 0.55 AU & 0.40-22.9 h\\  \hline
        11-02-2021  & Yes (Solar Flare) & Yes& 0.56 AU & 0.41-23.3 h\\  \hline
        01-12-2022  & No  & Yes & 0.76 AU & 0.56-31.7 h\\  \hline
        02-18-2022  & Yes (High-Speed Stream)& No & 0.32 AU & 0.22-13.3 h\\  \hline
        03-02-2022  & No & Yes & 0.24 AU & 0.18-10.0 h \\  \hline
        03-23-2022  & Yes (High-Speed Stream) & Yes& 0.64 AU & 0.47-26.7 h\\  \hline
        03-31-2022  & Yes (Solar Flare) & Yes& 0.71 AU & 0.53-29.6 h\\  \hline
        04-19-2022  & Yes (Solar Flare)& No  & 0.76 AU &0.56-31.7 h \\  \hline
        05-14-2022  & Yes (High-Speed Stream) & Yes & 0.56 AU & 0.41-23.3 h\\  \hline
        05-20-2022  & Yes (Solar Flare)& Yes & 0.45 AU & 0.34-18.8 h\\  \hline
        06-02-2022  & No  & Yes & 0.75 AU & 0.55-31.3 h \\  \hline
        07-09-2022  & Yes(Solar Flare) & Yes & 0.74 AU &0.54-30.8 h\\  \hline
        07-11-2022  & Yes (Solar Flare)& Yes & 0.75 AU & 0.55-31.3 h \\  \hline
        07-13-2022  & Yes (Solar Flare) & Yes & 0.75 AU & 0.55-31.3 h \\  \hline
        07-18-2022  & Yes (CME)& Yes & 0.76 AU & 0.56-31.7 h \\  \hline
        07-23-2022  & No & Yes & 0.76 AU & 0.56-31.7 h \\  \hline
        08-16-2022  & Yes (Solar Flare)& Yes & 0.58 AU & 0.42-24.2 h\\  \hline
        08-19-2022  & Yes (CME)& Yes & 0.53 AU & 0.38-22.1 h \\  \hline
        08-26-2022  & Yes (Solar Flare)& No & 0.41 AU & 0.34-17.1 h \\  \hline
        08-28-2022  & Yes (Solar Flare)& No & 0.36 AU & 0.27-15 h\\  \hline
        \hline\hline
    \end{tabular}
    \caption{ Index of alpha spike events and Associated CMEs, high speed streams, solar flares, and switchbacks at the relevant distance from the Sun. The rightmost column dictates the range of time for alpha particles in EPI-Lo energy ranges to reach PSP at that particular distance from the Sun.
             }
             \label{tab:table1}
\end{table}   
\begin{table}[htbp]
\centering
\begin{tabular}{|l|c|c|c|c|c|c|}\hline\hline
$\alpha$ spike & Associated with Solar Activity? & Associated $\alpha$-H SB? &Dist. from Sun & $\Delta$t for $\alpha$s from Sun \\ \hline\hline
        09-25-2022  & Yes (High-Speed Stream)& Yes & 0.57 AU & 0.42-23.8 h\\  \hline
        10-18-2022  & Yes (Solar Flare)  & Yes & 0.75 AU & 0.55-31.3 h\\ \hline
        12-12-2022  & No  & No & 0.08 AU & 0.06-3.4 h\\  \hline
        12-25-2022  & Yes (Solar Flare)  & Yes & 0.47 AU & 0.36-19.6 h\\ \hline
        01-04-2023  & Yes (CME)  & Yes & 0.62 AU & 0.47-25.8 h\\ \hline
        02-19-2023  & No & Yes & 0.65 AU & 0.36-27.1 h\\ \hline
        02-26-2023  & Yes (High-Speed Stream) & No & 0.58 AU & 0.44-24.2 h\\ \hline
        03-10-2023  & Yes (Solar Flare)  & No & 0.33 AU & 0.25-13.7 h\\ \hline
        03-13-2023  & Yes (CME)  & No & 0.24 AU & 0.18-10.0 h\\ \hline
        03-21-2023  & Yes (High-Speed Stream)  & No & 0.19 AU & 0.15-7.9 h\\ \hline
        04-23-2023  & Yes (CME)  & Yes & 0.73 AU & 0.56-30.5 h\\ \hline
        04-30-2023  & Yes (Solar Flare)  & No & 0.756 AU & 0.58-31.5 h\\ \hline
        05-12-2023  & Yes (Solar Flare)  & Yes & 0.75 AU & 0.57-31.3 h\\ \hline
        05-18-2023  & Yes (CME)  & Yes & 0.73 AU & 0.56-30.5 h\\ \hline
        06-21-2023  & Yes (Solar Flare)  & No & 0.06 AU & 0.47-25.8 h\\ \hline
        07-11-2023  & Yes (Solar Flare)  & Yes & 0.56 AU & 0.56-30.5 h\\ \hline
        07-19-2023  & Yes (Solar Flare)  & Yes & 0.66 AU & 0.51-27.5 h\\ \hline
        08-05-2023  & Yes (Solar Flare)  & No &  0.76 AU & 0.58-31.7 h \\ \hline
        09-01-2023  & Yes (CME)  & No & 0.66 AU & 0.51-27.5 h\\ \hline
        09-17-2023  & Yes (CME)  & No & 0.62 AU & 0.32-17.5 h\\ \hline
        09-23-2023  & Yes (Solar Flare)  & No & 0.27 AU & 0.20-11.3 h\\ \hline
        09-25-2023  & No  & No & 0.17 AU & 0.13-7.09 h\\ \hline
        09-26-2023  & No & No & 0.15 AU & 0.11-6.3 h\\ \hline
        09-27-2023  & Yes(High-Speed Stream)  & No & 0.10 AU & 0.08-4.17 h\\ \hline
        10-28-2023  & Yes (High-Speed Stream) & No & 0.685 AU & 0.52-28.6 h\\ \hline
        11-04-2023  & Yes (Solar Flare)  & Yes & 0.73 AU & 0.56-30.5 h\\ \hline
        01-04-2024  & Yes (Solar Flare)  & Yes & 0.26 AU & 0.20-10.8 h\\ \hline
        01-11-2024  & Yes (Solar Flare)  & Yes & 0.44 AU & 0.34-18.3 h\\ \hline
        01-23-2024  & Yes (Solar Flare)  & Yes & 0.64 AU & 0.49-26.7 h\\ \hline
        02-08-2024  & Yes (Solar Flare)  & No & 0.74 AU & 0.54-30.8 h \\ \hline
        02-10-2024  & Yes (Solar Flare)  & No & 0.74 AU & 0.54-30.8 h \\ \hline
        02-16-2024  & Yes (Solar Flare)  & Yes & 0.74 AU & 0.54-30.8 h \\ \hline
        02-18-2024  & Yes (Solar Flare)  & No &  0.74 AU & 0.54-30.8 h\\ \hline
        03-12-2024  & Yes (CME)  & No & 0.54 AU & 0.41-22.5 h\\ \hline
        03-17-2024  & Yes (Solar Flare)  & No & 0.46 AU & 0.35-19.2 h\\ \hline
        03-21-2024  & Yes (Solar Flare)  & No & 0.36 AU & 0.27-15.0 h\\ \hline
        03-26-2024  & Yes (Solar Flare)  & No & 0.25 AU & 0.19-10.4 h\\ \hline
        \hline\hline
    \end{tabular}
    \caption{  Part 2 of index of alpha spike events and Associated CMEs, high speed streams, solar flares, and switchbacks at the relevant distance from the Sun. The rightmost column dictates the range of time for alpha particles in EPI-Lo energy ranges to reach PSP at that particular distance from the Sun. }
             \label{tab:table1B}
\end{table}  

%

\begin{table}[htbp]
    \centering
    \begin{tabular}{|l|c|c|c|c|c|c|c|c|c|c|}                                    \hline\hline
        $\alpha$-H SB & $\Delta$ T & $\Delta$ Vr & $\Delta$ Vn & $\Delta$ Vt 
                   & $\Delta$ n  & SB dur  & $\Delta$ V dur  & $\Delta$t $\alpha$ spike & FA Currents?& Mirrors?\\ 
                   &  (\%) &  (\%) &  (\%) &  (\%) &  (\%) &  (min) &   (min)  &(h) & &  \\ \hline\hline   
        11-11-2018    & +20 & -3.8  & -28.6 & +67 & -50 & 90  & 90 & +12.5 & N/A & N/A\\ \hline
        05-29-2020     & -44.4  & -10   & -33.3  & -27.3  & -37.5 & 90  & 120 & +12 & No & No \\ \hline
        11-28-2020     & -42.9  & +200  &  0     &  +50   & -250  & 100 & 150& -45  & No & No \\ \hline
        11-29-2020     & -25    & -87.5 & +75    & -42.9  & +75   & 50  & 90& -20  & No & No\\ \hline
        07-01-2021     & +4.1   & -5.9  & -28.6  & +11.1  & +10.5 & 30  & 10& + 5 & No & No  \\\hline
        08-30-2021     &  N/A   &  N/A  &  N/A   &  N/A   & N/A   & 30  & N/A& +41 & No & No \\ \hline
        11-02-2021     &  N/A   &  N/A  &  N/A   &  N/A   & N/A   & 120 & N/A& -7 & No & No \\ \hline
        01-10-2022     &  N/A   &  N/A  &  N/A   &  N/A   & N/A   & 210 & N/A & -48 & No &No \\ \hline
        03-01-2022     & +42.8  & +7.7  & +181.8 & -118.2 & -66.7 & 50  & 50 & -10 & No & No\\ \hline
        03-22-2022     & +10    & +3.4  & +12    & -11.1  & -50   & 10   & 10  & -28 & No & No \\ \hline
        03-31-2022     &  N/A   &  N/A  &  N/A   &  N/A   & N/A   & 19  & N/A& -17& No & No\\ \hline
        05-12-2022     & -28    & +7.1  & +100   & -50    & -42.8 & 130 & 130& -43 & No & No\\ \hline
        05-13-2022     & +4.2   & -6.4  & -20    & +25    & -33.3 & 34   & 34 & -37 & No & No \\ \hline
        05-14-2022     & -60    & -6.3  & +60    & -25    & +175  & 60  & 80
        & -9 & Yes & No  \\ \hline
        05-19-2022     & -10    & +6.3  & +50    & -60    & -27.3 & 120 & 180& -30 & No & No \\ \hline
        05-20-2022 (1) & +22.2  & -2.9  & -166.7 & +120   & +140  & 60  & 60  &+3 & No & No \\ \hline
        05-20-2022 (2) & -21.9  & +5.9  & -66.7  & +16.7  & -80   & 60  & 70  &+9 & N/A & N/A\\ \hline
        06-04-2022 (1) & +120   & +40   & -1100  & +166.7 & -93.3 & 30  & 30 & +39 & No & No \\ \hline
        06-04-2022 (2) & +140   & +80   & -1000  & +120   & -98.3 & 30  & 30  & +42 & N/A & N/A\\ \hline
        06-04-2022 (3) & -21    & -19.2 & -200   & +100   & +150  & 60  & 60  & +48 & No & No\\ \hline
        07-09-2022     & +7.7   & -5.3  & -200   & +28.6  & +150  & 50  & 60 &-1 & N/A &N/A \\ \hline
        07-10-2022 (1) & +9.1   &  0    & -60    & +50    & +166.7& 15   & 15  &+3 & N/A & N/A \\ \hline
        07-10-2022 (2) & -16.7  & -12.5 &  0     & 0      & -20   & 15  & 15  & +3 & N/A & N/A\\ \hline
        07-10-2022 (3) & +66.7  & +11.4 & -66.7  & 0      & -44.4 & 30  & 30 & +9 & N/A & N/A \\ \hline
        07-11-2022 (1) & +22.6  & +18.8 & +14.3  & -200   & -66.7 & 50  & 50  &-6 & N/A & N/A\\ \hline
        07-11-2022 (2) & -83.3  & +9.4  & +100   & -166.7 &  +140 & 50  & 50 & +1 & N/A &N/A \\ \hline
        07-12-2022     & +40    & -3    & -25    & +28.6  & +83.3 & 40  & 30  & +29 & N/A & N/A \\ \hline
        07-13-2022     & +20    & +11.4 & +100   & -28.6  & +400  & 45  & 45  & +20 & N/A &N/A\\ \hline
        07-14-2022 (1) & +23.8  & -4.9  &  0     & -28.6  &  +40  & 25  & 25 & +44 & N/A &N/A \\ \hline 
        07-14-2022 (2) & +35.3  & +5.9  &  +60   & -14.3  & +120  & 30  & 90 &+34 & N/A &N/A \\ \hline 
        07-14-2022 (3) & -23    & +8.6  & +33.3  & -28.6  & -25   & 50  & 50  &+37 & N/A & N/A\\ \hline 
        07-17-2022     & -70.8  & -12.8 & -66.6  & +55.6  & +87.5 & 60  & 60  &- 29 & N/A &N/A\\ \hline 
        07-23-2022 (1) & -12.5  & -6.9  & +66.7  & -100   & +40   & 60  & 60 & -8 & N/A & N/A\\ \hline
        07-23-2022 (2) & -30.8  & -6.9  & +33.3  & 0      & +175  & 60  & 60 & +9 & N/A &N/A\\ \hline
        08-17-2022     & +50    & -4.3  & +66.7  & -200   & -64.7 & 100 & 100 & +26 & No & No\\ \hline
        08-19-2022     & +25    & +3.6  & -800   & +35.7  & -90.9 & 120 & 120 & +29 & No & No\\ \hline
        08-20-2022     & +254.8 & +4.7  & -400   & +50    & -95   & 40  & 40 & + 48 & No & No\\ \hline
        \hline
    \end{tabular}
    \caption{ Index of alpha-heavy Switchback Events and the percent change of the 
             associated plasma from the surrounding solar wind of the associated 
             solar wind properties: temperature, density, and velocity in RTN 
             coordinates plus the duration of the switchback contrasted with the 
             perturbed plasma velocity duration, the presence of field-aligned currents, and the presence of mirroring. Non available data is marked N/A.
             }
            \label{tab:table2}
\end{table}         
\begin{table}[htbp]
    \centering
    \begin{tabular}{|l|c|c|c|c|c|c|c|c|c|c|}                                    \hline\hline
        $\alpha$-H SB & $\Delta$ T & $\Delta$ Vr & $\Delta$ Vn & $\Delta$ Vt 
                   & $\Delta$ n  & SB dur  & $\Delta$ V dur  & $\Delta$t $\alpha$ spike & FA Currents?& Mirrors?\\ 
                   &  (\%) &  (\%) &  (\%) &  (\%) &  (\%) &  (min) &   (min)  &(h) & &  \\ \hline\hline   
        09-25-2022     & --60   & -13.3 & -16.7  & +14.3  & +200  & 30  & 30  & +21 & No  &No\\ \hline
        09-26-2022     & +60    & -6.5  & +200   & -50    & -97.5 & 20  & 20 & + 26 & No &No \\ \hline
        10-17-2022     & +50.9  & +6.3  & +33.3  & -7.7   & -75   & 60  & 60 & +3 & No & No  \\ \hline 
        10-19-2022     & -23.8  & -5.8  & +50    & +9     & +33.3 & 30  & 30 & +34 &No & No \\ \hline
        12-25-2022     & +300  & +600  & -200    & +9     & -100 & 40  & 40 & +10 &No & No\\ \hline
        01-04-2023     & -20  & +24 & -1050  & +20  & +116.7 & 20  & 60 & -3 &No & No\\ \hline
        02-19-2023     & -25  & +10 & -38  & +60  & +100 & 90  & 120 & -12 &No & No\\ \hline
        04-22-2023     & -25  &-8.6 & +25  & -40  & +200 & 20  & 40 & -24 &No & No\\ \hline
        05-10-2023     & +59.1  & +10.5 & +66.7  & -700  & -50 & 120  & 120 & -28 &No & No\\ \hline
        05-16-2023     & +66.7  & +400 & +50  & -700  & -80 & 30  & 30 & -40 &No & No\\ \hline
        07-12-2023     & -25  & +7.1 & -100  & -200  & +300 & 30  & 40 & +15 &No & No\\ \hline
        07-18-2023     & -40  & +300 & -100  & -120  & +200 & 120  & 180 & -30 &No & No\\ \hline
        11-05-2023     & -50  & +12.5 & +100  & -120  & +900 & 120 & 140 & +25 &No & No\\ \hline
        01-03-2024     &+50  & +7.15 & -300  & -400  & -80 & 60 & 60 & -5 &No & No\\ \hline
        01-12-2024     & -75  & +2.7 & -115.7  & -162.5  & +900 & 60 & 100 & +27 &No & No\\ \hline
        01-21-2024     & +200  & -8.3 & +66.7  & +25 & -80 & 120 & 120 & -41 & No & No\\ \hline
        02-21-2024     & 0  & +100 & -100  & +100 & N/A & 70 & 70 & -36 & No & No\\ \hline
        \hline
    \end{tabular}
    \caption{ Part 2 of the index of alpha-heavy Switchback Events and the percent change of the 
             associated plasma from the surrounding solar wind of the associated 
             solar wind properties: temperature, density, and velocity in RTN 
             coordinates plus the duration of the switchback contrasted with the 
             perturbed plasma velocity duration, the presence of field-aligned currents, and the presence of mirroring. Non available data is marked N/A.  
             }
            \label{tab:table2B}
\end{table}         

\begin{table}[htbp]
\centering
\begin{tabular}{|l|c|c|c|}\hline\hline
Month & \# of $\alpha$ Spikes  & \# of Regular Switchbacks&  \# of $\alpha$-H SBs \\ \hline\hline
        10-2018& 0  & 65 & 0  \\  \hline
        11-2018& 1 & 63  & 2 \\  \hline
        12-2018& 0 & 37  & 2 \\  \hline
        06-2022& 1 & 68 & 9 \\  \hline
        07-2022& 5 & 69 & 21\\  \hline
        08-2022& 4 & 59 & 10 \\  \hline\hline
    \end{tabular}
    \caption{ Average number of observed events per week of alpha-heavy switchbacks and non-alpha-heavy switchbacks during a 3~month period of low solar activity (October-December 2018) and a 3~month period of high solar activity (June-August 2022).}
        
             \label{tab:table3}
\end{table}         
\begin{table}[htbp]
    \centering
    \begin{tabular}{|l|c|c|c|c|c|c|}                                    \hline\hline
        $\alpha$-H SB & Br fluct. & Corr. with Vr& Dist. from Sun & $\Delta t$ for $\alpha$s from Sun & Pass W\`{a}len Test & $>$ err for FMSW \\ \hline\hline                   
        11-11-2018     & Yes & Yes (-) & 0.24 AU & 0.17-10.6 h & No & No \\ \hline
        05-29-2020     & Yes & Yes & 0.34 AU & 0.25-14.2 h & No & No  \\ \hline
        11-28-2020     & Yes & Yes & 0.81 AU & 0.61-33.7 h & No & No \\ \hline
        11-29-2020     & Yes & Yes & 0.81 AU & 0.61-33.7 h & No & No \\ \hline
        07-01-2021     & Yes & Yes & 0.75 AU & 0.55-31.3 h & No & No \\\hline
        08-30-2021     & N/A & N/A & 0.58 AU & 0.42-23.8 h & No & No \\ \hline
        11-02-2021     & N/A & N/A & 0.56 AU & 0.41-23.3 h & No& No \\ \hline
        01-10-2022     & N/A & N/A & 0.76 AU& 0.56-31.7 h& No & No \\ \hline
        03-01-2022     & Yes & Yes (-)& 0.22 AU & 0.16-9.2 h & No & No \\ \hline
        03-22-2022     & Yes & Yes (-)& 0.63 AU &0.46-26.3 h & No & No \\ \hline
        03-31-2022     & N/A & N/A & 0.71 AU & 0.53-29.6 h & No & No \\ \hline
        05-12-2022     & Yes & Yes & 0.57 AU & 0.42-23.8 h & No & No  \\ \hline
        05-13-2022     & Yes & Yes & 0.57 AU & 0.42-23.8 h & No & No \\ \hline
        05-14-2022     & Yes & Yes & 0.56 AU & 0.41-23.3 h & No & No   \\ \hline
        05-19-2022     & Yes & Yes (-) & 0.47 AU & 0.36-19.6 h & No & No \\ \hline
        05-20-2022 (1) & Yes & Yes (-) & 0.45 AU & 0.34-18.8 h & No & No   \\ \hline
        05-20-2022 (2) & Yes & Yes (-) & 0.44 AU & 0.33-18.3 h & No & No \\ \hline
        06-04-2022 (1) & Yes & Yes & 0.13 AU & 0.1-5.4 h & No & No \\ \hline
        06-04-2022 (2) & Yes & Yes & 0.14 AU & 0.11-5.8 h & No & No  \\ \hline
        06-04-2022 (3) & Yes & Yes (-) & 0.16 AU & 0.12-6.7 h & No & No  \\ \hline
        07-09-2022     & Yes & Yes & 0.74 AU & 0.54-30.8 h & No & No   \\ \hline
        07-10-2022 (1) & Yes & No & 0.75 AU & 0.55-31.3 h & No & No   \\ \hline
        07-10-2022 (2) & Yes & Yes& 0.75 AU & 0.55-31.3 h & No & No   \\ \hline
        07-10-2022 (3) & Yes & Yes& 0.75 AU & 0.55-31.3 h & No & No \\ \hline
        07-11-2022 (1) & Yes & Yes (-)& 0.74 AU & 0.54-30.8 h & No & No \\ \hline
        07-11-2022 (2) & Yes & Yes & 0.75 AU & 0.55-31.3 h & No & No \\ \hline
        07-12-2022     & Yes & Yes (-)& 0.75 AU &0.55-31.3 h & No & No \\ \hline
        07-13-2022     & Yes & Yes (-)& 0.75 AU & 0.55-31.3 h & No & No  \\ \hline
        07-14-2022 (1) & Yes & Yes (-)& 0.75 AU & 0.55-31.3 h & No& No \\ \hline 
        07-14-2022 (2) & Yes & Yes (-)& 0.75 AU & 0.55-31.3 h & No & No  \\ \hline 
        07-14-2022 (3) & Yes & Yes (-)& 0.75 AU & 0.55-31.3 h & No & No \\ \hline 
        07-17-2022     & Yes & Yes (-)& 0.76 AU & 0.56-31.7 h & No & No \\ \hline 
        07-23-2022 (1) & Yes & Yes & 0.76 AU & 0.56-31.7 h & No & No \\ \hline
        07-23-2022 (2) & Yes & Yes & 0.76 AU & 0.56-31.7 h & No &No \\ \hline
        08-17-2022     & Yes & Yes & 0.57 AU & 0.42-23.8 h & No & No  \\ \hline
        08-19-2022     & Yes & Yes & 0.53 AU & 0.38-22.1 h & No & No  \\ \hline
        08-20-2022     & Yes & Yes & 0.52 AU & 0.37-21.7 h & No & No  \\ \hline
        \hline
    \end{tabular}
    \caption{Index of alpha-heavy Switchback Events and associated properties. These properties include whether there are fluctuations in the radial magnetic field, whether those magnetic field fluctuations correlate or anticorrelate with the fluctuations in the radial solar wind velocity component, the distance from the Sun of the alpha-heavy switchback, the time range it would take alpha particles in EPI-Lo energy ranges to reach $PSP$ at that distance from the Sun, whether the switchbacks pass the W\`{a}len test, and whether the arrival times for each switchback are outside the error for a fast magnetosonic wave.
    Non available data is marked N/A. 
             }
            \label{tab:table4}
\end{table}         
\begin{table}[htbp]
    \centering
    \begin{tabular}{|l|c|c|c|c|c|c|}                                    \hline\hline
        $\alpha$-H SB & Br fluct. & Corr. with Vr& Dist. from Sun & $\Delta t$ for $\alpha$s from Sun & Pass W\`{a}len Test & $>$ err for FMSW \\ \hline\hline                   
        09-25-2022     & Yes & Yes (-)&0.57 AU & 0.42-23.8 h & No & No \\ \hline
        09-26-2022     & Yes & Yes  & 0.57 AU & 0.42-23.8 h & No & No \\ \hline
        10-17-2022     & Yes & Yes (-)& 0.75 AU &0.55-31.3 h & No & No \\ \hline 
        10-19-2022     & Yes & Yes & 0.75 AU & 0.55-31.3 h & No & No\\ \hline
        12-25-2022     & Yes & Yes & 0.47 AU &0.36-19.6 h & No & No \\ \hline 
        01-04-2023     & Yes & Yes & 0.62 AU & 0.47-25.8 h & No & No\\ \hline
        02-19-2023     & Yes & Yes &0.65 AU & 0.36-27.1 h& No & No\\ \hline
        04-22-2023     & Yes & Yes (-) & 0.727 AU & 0.55-30.3 h& No & No\\ \hline
        05-10-2023     & Yes & Yes (+/-) & 0.75 AU & 0.57-31.3 h& No & No\\ \hline
        05-16-2023     & Yes & Yes (-) & 0.73 AU & 0.56-30.5 h& No & No\\ \hline
        07-12-2023     & Yes & Yes & 0.58 AU & 0.44-24.2 h & No & No\\ \hline
        07-18-2023     & Yes & Yes (-) & 0.65 AU & 0.50-27.1 h & No & No\\ \hline
        11-05-2023     & Yes & Yes (-) &  0.73 AU & 0.56-30.5 h & No & No\\ \hline
        01-03-2024     & Yes & Yes (-) &  0.25 AU & 0.19-10.4 h & No & No\\ \hline
        01-12-2024     & Yes & Yes  &  0.48 AU & 0.37-20.0 h & No & No\\ \hline
        01-21-2024     & Yes & Yes  &  0.625 AU & 0.48-26.1 h & No & No\\ \hline
        02-15-2024     & Yes & Yes  & 0.74 AU & 0.54-30.8 h  & No & No\\ \hline
        \hline
    \end{tabular}
    \caption{Part 2 of the index of alpha-heavy Switchback Events and associated properties. These properties include whether there are fluctuations in the radial magnetic field, whether those magnetic field fluctuations correlate or anticorrelate with the fluctuations in the radial solar wind velocity component, the distance from the Sun of the alpha-heavy switchback, the time range it would take alpha particles in EPI-Lo energy ranges to reach $PSP$ at that distance from the Sun, whether the switchbacks pass the W\`{a}len test, and whether the arrival times for each switchback are outside the error for a fast magnetosonic wave.
    Non available data is marked N/A.}

            \label{tab:table4b}
\end{table}    

\end{document}